# MONITORING AND OPTIMIZATION OF ENERGY CONSUMPTION OF BASE TRANSCEIVER STATIONS

Antonio Spagnuolo[1*], Antonio Petraglia[2], Carmela Vetromile[1], Roberto Formosi[1], Carmine Lubritto[1]
[1]Department of Environmental Science and Technology (DiSTABiF), Second University of Naples
Via Vivaldi, 43, I - 81100 Caserta – ITALY
[2]Department of Mathematics and Physics (DMF), Second University of Naples
Viale A. Lincoln, 54, I - 81100 Caserta – ITALY

*Corresponding author: antonio.spagnuolo@unina2.it

**Highlights**

- Energy consumption and environmental parameters of a base transceiver system have been monitored.
- Energy consumption is related to the air conditioning functions and to the load of telephone traffic.
- Energy saving can be obtained by careful choice of cooling parameters and by turn off BS transceivers
- Energy saving parameters can be estimated by a simulation Monte Carlo method.

**Abstract**

The growth and development of the mobile phone network has led to an increased demand for energy by the telecommunications sector, with a noticeable impact on the environment.
Monitoring of energy consumption is a great tool for understanding how to better manage this consumption and find the best strategy to adopt in order to maximize reduction of unnecessary usage of electricity. This paper reports on a monitoring campaign performed on six Base Transceiver Stations (BSs) located central Italy, with different technology, typology and technical characteristics.
The study focuses on monitoring energy consumption and environmental parameters (temperature, noise, and global radiation), linking energy consumption with the load of telephone traffic and with the air conditioning functions used to cool the transmission equipment. Moreover, using experimental data collected, it is shown, with a Monte Carlo simulation based on power saving features, how the BS monitored could save energy.

**Abbreviations**

| | |
|---|---|
| BS | Base Transceiver Station |
| TRX | Transceiver |
| GSM | Global System for Mobile Communications |
| DCS | Digital Cellular System |
| UMTS | Universal Mobile Telecommunications System |
| MRFU | Multiple Radio Filter Unit |
| DRFU | Double Radio Filter Unit |
| BCCH | Broadcast Control Channel |
| SDCCH | Standalone Dedicated Control Channel |
| FC | Free Cooling System |
| CDZ | Air Conditioning System |
| ITU | International Telecommunication Union |
| ETSI | European Telecommunications Standards Institute |

| TA | Timing Advance |
|---|---|
| MR | Measurement Report |

## 1. Introduction

Telecommunications is one of the sectors where the continuous growth in demand for mobile services and the parallel technological development go hand in hand with regards to energy consumption; it suffice to think that information and communications technology (ICT) is accountable for consumption of about 3% of the world's total electrical energy. By the end of 2030, it is expected that this figure will grow to 1,700 TWh [1]

For this reason, the issue of BSs energy management is fundamental for sustainable development of the sector [2].

Previous literature proposes a number of solutions for efficiency and / or energy savings.

The first set of energy-saving suggestions focused on transmission functions. For example, it was proposed to act on the range of action of the plant (cell zooming) [3], on the high-efficiency power amplifiers [4], on the technological improvement of radio frequency units and smart antenna technology [5], by setting the BSs to stand-by mode [5][6][7][8], in order to correlate network energy consumption and telephonic traffic load [9][10].

Furthermore, energy consumption can also be reduced by acting on the air-conditioning systems, both via a correct setting of the apparatus system control [11][12] and innovative conditioning systems [13][14][15].

To better implements the energy saving actions and, consequently, improve the environmental impact of the cellular networks, it is necessary to accurately monitor the energy consumed by a BS.

The aim of this paper is the study and analysis of energy consumption of a BS, and related environmental parameters, through a monitoring campaign aimed at getting a clear and complete picture of the consumption dynamics of a BS, and the variables that exert more influence on its performance.

## 2. Characteristics of the base transceiver stations studied

The monitoring involved six BSs of the telephone service provider Wind, located in the municipalities of Frosinone, Sora (FR), Cassino (FR), and Pontecorvo (FR) in center Italy. Table 1 reports the technical characteristics of the BS studied: site name, code and size, BS typology (Shelter, Outdoor, and Room), BS technology (GSM/DCS/UMTS), configuration for each technology (number of TRXs per sector), radio unit type (DRFU, MRFU), number of timeslots for voice, data, BCCH, SDCCH.

| SITE CODE | | | FR001 | | | | | SITE CODE | | | FR009 | | | | |
|---|---|---|---|---|---|---|---|---|---|---|---|---|---|---|---|
| Site name | | | FROSINONE CENTRO | | | | | Site name | | | PAREDA | | | | |
| Station type | | | SHELTER | | | | | Station type | | | SHELTER | | | | |
| Size | | | 7 m$^2$ | | | | | Size | | | 7 m$^2$ | | | | |
| Sampling period | | | 24/09/2013 – 02/10/2013 // 04/12/2013 – 11/12/2013 | | | | | Sampling period | | | 17/09/2013 – 24/09/2013 | | | | |
| GSM Config. | | | 2-2-2 | | | | | GSM Config. | | | 2-2-2 | | | | |
| DCS Config. | | | 4-4-6 | | | | | DCS Config. | | | 4-4-4 | | | | |
| UMTS Config. | | | 1-1-1 | | | | | UMTS Config. | | | 1-1-1 | | | | |
| Sector | Radio unit | Type | TRX | Dedicated Voice timeslots | Dedicated data timeslot | BCCH | SDCCH | Sector | Radio unit | Type | TRX | Dedicated Voice timeslots | Dedicated data timeslot | BCCH | SDCCH |
| FR001D1 | 1 | MRFU | 4 | 28 | 1 | 1 | 2 | FR009D1 | 1 | MRFU | 4 | 28 | 1 | 1 | 2 |
| FR001D2 | 1 | MRFU | 4 | 28 | 1 | 1 | 2 | FR009D2 | 1 | MRFU | 4 | 28 | 1 | 1 | 2 |
| FR001D3 | 2 | MRFU | 6 | 43 | 1 | 1 | 3 | FR009D3 | 1 | MRFU | 4 | 28 | 1 | 1 | 2 |
| FR001G1 | 2 | MRFU | 2 | 12 | 1 | 1 | 2 | FR009G1 | 1 | MRFU | 2 | 12 | 1 | 1 | 2 |
| FR001G2 | 2 | MRFU | 2 | 12 | 1 | 1 | 2 | FR009G2 | 1 | MRFU | 2 | 12 | 1 | 1 | 2 |
| FR001G3 | 2 | MRFU | 2 | 13 | 1 | 1 | 2 | FR009G3 | 1 | MRFU | 2 | 13 | 1 | 1 | 1 |
| SITE CODE | | | FR039 | | | | | SITE CODE | | | FR049 | | | | |
| Site name | | | SORA CENTRO | | | | | Site name | | | PONTECORVO | | | | |
| Station type | | | ROOM | | | | | Station type | | | ROOM | | | | |

| Size | 15 m² | | | | | | Size | 10 m² | | | | | |
|---|---|---|---|---|---|---|---|---|---|---|---|---|---|
| Sampling period | 22/10/2013 – 29/10/2013 | | | | | | Sampling period | 30/10/2013 – 06/11/2013 | | | | | |
| GSM Config. | 2-2-2 | | | | | | GSM Config. | 2-2-2 | | | | | |
| DCS Config. | 2-2-2 | | | | | | DCS Config. | 2-2-2 | | | | | |
| UMTS Config. | 1-1-1 | | | | | | UMTS Config. | 1-1-1 | | | | | |
| Sector | Radio Unit | Type | TRX | Dedicated Voice timeslots | Dedicated data timeslot | BCCH | SDCCH | Sector | Radio unit | Type | TRX | Dedicated Voice timeslots | Dedicated data timeslot | BCCH | SDCCH |
| FR039D1 | 1 | DRFU | 2 | 13 | 1 | 1 | 1 | FR049D1 | 1 | MRFU | 3 | 20 | 1 | 1 | 2 |
| FR039D2 | 1 | DRFU | 2 | 13 | 1 | 1 | 1 | FR049D2 | 1 | DRFU | 2 | 13 | 1 | 1 | 1 |
| FR039D3 | 1 | DRFU | 2 | 13 | 1 | 1 | 1 | FR049D3 | 1 | DRFU | 2 | 12 | 1 | 1 | 2 |
| FR039G1 | 2 | MRFU | 2 | 13 | 1 | 1 | 1 | FR049G1 | 1 | MRFU | 3 | 20 | 1 | 1 | 2 |
| FR039G2 | 2 | MRFU | 2 | 13 | 1 | 1 | 1 | FR049G2 | 1 | MRFU | 2 | 12 | 1 | 1 | 2 |
| FR039G3 | 2 | MRFU | 2 | 13 | 1 | 1 | 1 | FR049G3 | 1 | MRFU | 2 | 13 | 1 | 1 | 1 |
| SITE CODE | FR005 | | | | | | SITE CODE | FR011 | | | | | |
| Site name | FICUCCIA | | | | | | Site name | SAN GIOVANNI | | | | | |
| Station type | OUTDOOR | | | | | | Station type | OUTDOOR | | | | | |
| Size | Only cabinet | | | | | | Size | Only cabinet | | | | | |
| Sampling period | 02/10/2013 – 09/10/2013 | | | | | | Sampling period | 08/11/2013 – 15/11/2013 | | | | | |
| GSM Config. | 2-2-2 | | | | | | GSM Config. | 4-2-0 | | | | | |
| DCS Config. | 3-2-4 | | | | | | DCS Config. | 6-4-0 | | | | | |
| UMTS Config. | 1-1-1 | | | | | | UMTS Config. | 1-1-0 | | | | | |
| Sector | Radio Unit | Type | TRX | Dedicated Voice timeslots | Dedicated data timeslot | BCCH | SDCCH | Sector | Radio unit | Type | TRX | Dedicated Voice timeslots | Dedicated data timeslot | BCCH | SDCCH |
| FR005D1 | 1 | MRFU | 3 | 20 | 1 | 1 | 2 | FR011D1 | 2 | MRFU | 6 | 44 | 1 | 1 | 2 |
| FR005D2 | 1 | DRFU | 2 | 12 | 1 | 1 | 2 | FR011D2 | 2 | MRFU | 4 | 28 | 1 | 1 | 2 |
| FR005D3 | 1 | MRFU | 4 | 28 | 1 | 1 | 2 | FR011G1 | 1 | MRFU | 4 | 28 | 1 | 1 | 2 |
| FR005G1 | 1 | DRFU | 2 | 13 | 1 | 1 | 1 | FR011G2 | 1 | MRFU | 2 | 12 | 1 | 1 | 2 |
| FR005G2 | 2 | MRFU | 2 | 13 | 1 | 1 | 1 | | | | | | | | |
| FR005G3 | 2 | MRFU | 2 | 12 | 1 | 1 | 2 | | | | | | | | |

**Table 1. Technical characteristics of the stations.**

Regarding the typology, the characteristics of different BSs are indicated below:
**Shelter**: cabins made of aluminum and polyurethane foam containing the transmission equipment (housed in special boxes), air conditioning system and all that is needed for the correct functioning of the BS.
**Room**: buildings containing the same equipment as a shelter.
**Outdoor**: box containing the transmission equipment. Neither any kind of coverage, nor air conditioning systems are present.
The cooling of indoor air is required in the room and shelter because of the high temperatures of the indoor environment; this is due to the large amount of heat dissipated by the transmission systems and to the factor of solar radiations falling on the cabin (the latter is obviously also present in the case of outdoor BSs, though they are not provided with air-conditioning systems).
The methodologies usually used to decrease the temperature inside the shelter can be identified as:
- **Free cooling (FC):** cooling system that utilizes atmospheric air temperature to lower the temperature inside the structure: the air enters the room from the outside through a vent.

- **Air conditioner (CDZ)**

In the case study, for both types of BS (shelter, room), the free cooling system and the air conditioner are located on the same machine and are ON / OFF type. When free cooling is turned on, an extractor is activated, which stops the enclosure from overpressuring.

The on / off of the air conditioner and free cooling is controlled by a PLC (Programmable Logic Controller), with preset "ON" thresholds and associated hysteresis values. The latter represents the temperature drop needed (after reaching the "ON" threshold) so that the cooling machine turns itself off. Table 2 shows the thresholds of switch-on and the related hysteresis values for the Shelter and Room stations studied.

|  |  | SHELTER |  | ROOM |  |
|---|---|---|---|---|---|
|  |  | FR001 | FR009 | FR039 | FR049 |
| FC1 | THRESHOLD (°C) | 24 | 24 | 26 | 22 |
|  | HYSTERESIS | 1 | 1 | 2 | 1 |
| FC2 | THRESHOLD (°C) | 25 | 24 | 27 | 23 |
|  | HYSTERESIS | 2 | 1 | 2 | 1 |
| CDZ 1 | THRESHOLD (°C) | 27 | 27 | 28 | 27 |
|  | HYSTERESIS | 1 | 1 | 2 | 1 |
| CDZ 2 | THRESHOLD (°C) | 28 | 28 | 29 | 28 |
|  | HYSTERESIS | 1 | 1 | 2 | 1 |

**Table 2. Thresholds of switch-on and the related hysteresis values for the Shelter and Room stations.**

Over the ON / OFF thresholds presented in the Table 2, the PLC indicates activation of the free cooling if the outdoor temperature is lower than 24 °C, and if the difference between indoor and outdoor temperature is higher than 2 °C.

### 3. Materials and methods

In the monitoring campaign, the features of stations (shelters, room, outdoor) and the following operating parameters were considered separately:
- Energy consumption (kWh)
- Indoor Station Temperature (°C)
- Outdoor Station Temperature (°C)
- Noise (dBA)
- Global Radiation (W/m$^2$)

The measurements were acquired via:
- **LSI LASTEM - Elog:** a data-logger for environmental applications, to which the signals from the various sensors used for the measurements were fed. Specifically:
    **Pt100 probe:** it provides a temperature reading. Apart from the Pt100, the LSI LASTEM is equipped with its own inner thermocouples. This allowed us to read both the internal and external temperature of the station.
    **Phonometer Smart Sensor Model AR814**: noise was measured for better understanding of the dynamics of the ON / OFF status of the air conditioning equipment.
    **Radiometer:** the measurement of global radiation allows for control of environmental factors outside the station.
- **Wireless Monitor CM160:** data-logger that allows to record electricity consumption in real time. The monitor acquires data sent from a transmitter to which three amperometric clamps (one for each phase of the electrical panel of the station) are connected.
- **Thermometer CAR-889:** infrared thermometer with a laser indicator to measure the temperature of the individual devices.

The data collection period lasted one week (7 consecutive days) for each station, during which averages per minute for all measured variables were acquired. Therefore, in total, for each measurement, 10,080 measured values were recorded. Furthermore, the traffic data (hourly averages of traffic voice and data) of the stations was also retrieved from the operator, for the duration of the sampling.

## 4. Results and discussions

### 4.1 Analysis of total energy consumption

The first result relates to the average daily and yearly energy consumption of a BS, depending on its typology.
The measurement campaign was performed in early autumn, and these researchers consider it valid to simulate the annual average environmental conditions (ambient temperature, solar radiation and telephone traffic) for Central Italy; it is possible to calculate annual energy consumption from daily experimental results. Moreover, the figures on the annual energy consumption are consistent with those reported in [16], where the energy consumption of BSs were calculated from statistical data of annual consumption for 95 BSs on the Italian territory. In other words, the experimental conditions allow to consider annual data from daily or hourly results as representative of the system for the whole year.
From the analysis of the energy consumption of 6 BSs examined, reported in Table 3, it emerges that the average daily consumption of a BS ranges between 41 kWh and 117 kWh, depending on the type of BS. These result permits to calculate, considering that each BS is on for 365 days for year, an annual energy consumption of a BS ranges between 14,965 kWh and about 42,705 kWh: it is higher for shelter (daily average 108 kWh) and room (daily average 76.5 kWh), and lower for the outdoor type (daily average 53 kWh).
Starting from the average daily energy consumption, $CO_2$ emissions for the different BSs monitored have been calculated following the procedure defined by IPCC guidelines in [17] and values are reported in Table 3.

| Site ID | Type | Municipality | Average hourly consumption (kWh) | Average daily consumption (kWh) | Average annual consumption (kWh) | $CO_2$ Emissions (t/year) |
|---|---|---|---|---|---|---|
| FR001 | Shelter | FROSINONE | 4.9 | 117 | 42,705 | 20.55 |
| FR009 | Shelter | FROSINONE | 4.1 | 99 | 36,135 | 17.51 |
| FR039 | Room | SORA | 3.2 | 77 | 28,105 | 13.63 |
| FR049 | Room | PONTECORVO | 3.1 | 76 | 27,740 | 13.33 |
| FR005 | Outdoor | FROSINONE | 2.7 | 65 | 23,725 | 11.24 |
| FR011 | Outdoor | CASSINO | 1.7 | 41 | 14,965 | 7.16 |

**Table 3. Average hourly, daily, annual consumption and $CO_2$ emissions for each type station.**

These results are justified by the fact that BSs "Room" and "Shelter" require more energy, due to the air conditioning of the "building" that contains the devices needed for the operation of the entire plant.
It is possible to distinguish between the energy consumption of the cooling system, approximately 40 - 45 % of the total energy, and that of the transmission functions, which is about 55 % - 60 % of the total consumption of the system, confirming previous literature results [16].
Fig. 1, through the monitoring results, exemplifies the operating principle of free cooling and air conditioners for the period considered. In detail, in the Fig. 1 are shown: the internal temperature (blue line), the outside temperature (red line) and the set point of free cooling and air conditioners. Note that the free cooling are always ON at night and in the evening, when the internal temperature is always above the threshold of Table 2, while the air conditioner 1 is ON when the threshold is equal to 27 °C and OFF at the hysteresis value of 26 °C ((27-1) °C). During the day, the two free cooling are OFF because the outside temperature is above 24 °C, and then they resume normal operation at night, when the outdoor temperature returns below the threshold.

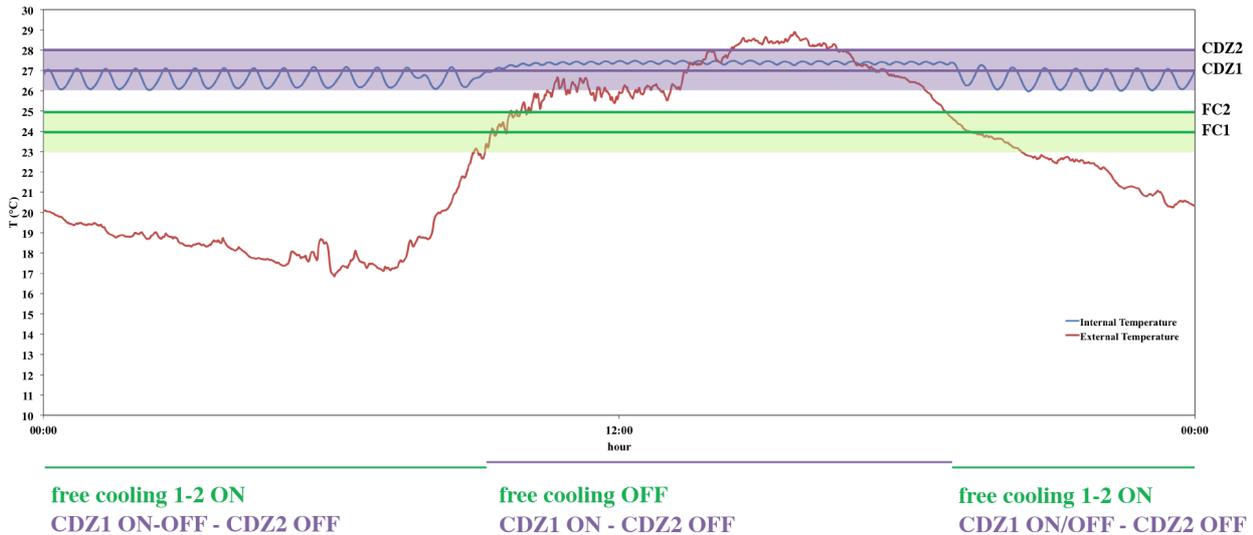

**Fig. 1. Operating principle of free cooling and conditioners.**

**4.2 Energy consumption versus traffic load**

To understand the characteristics of energy consumption of the BS, first the amount of traffic handled by the various BSs must be taken into account.

To study the behavior of energy consumption as a function of the load of telephone traffic, this paper considers the "outdoor" station in Table 1 (FR005). It was chosen because it does not require energy to power the air conditioning; hence, it is possible to easily correlate the energy consumption only to the transmission functions (i.e. traffic load).

A typical weekly traffic behavior is presented in Fig. 2, where the energy consumption related to the traffic load for the "outdoor" station FR005 is shown: the horizontal axis shows the sampling period, the ordinate axis on the left shows the energy consumption (kWh - red line), while the ordinate axis on the right shows the total telephone traffic (erl) (purple line), the voice traffic (green line) and the data traffic (yellow line)

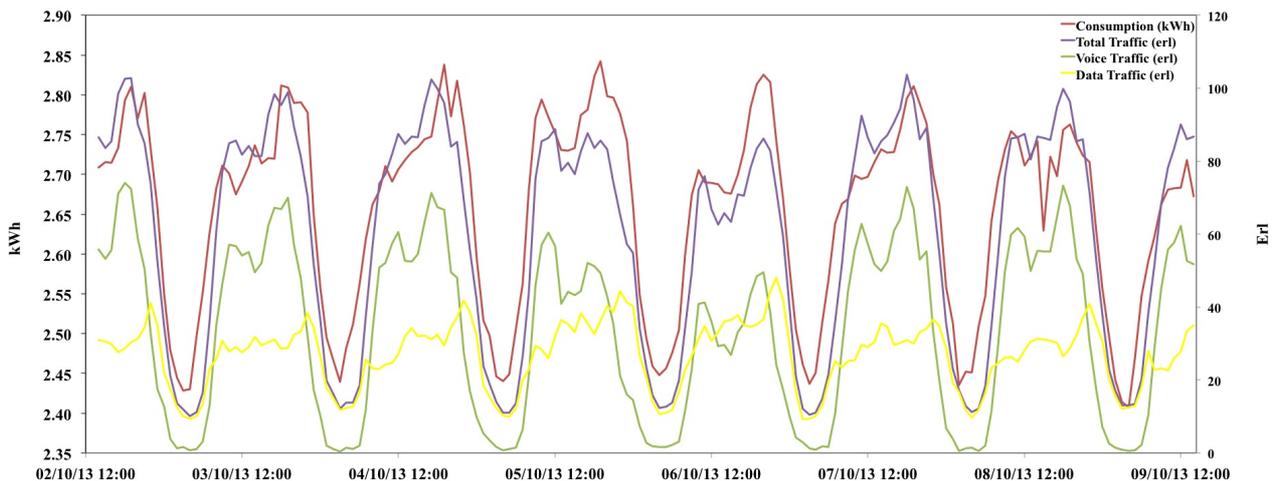

**Fig. 2. Energy consumption and traffic load for the outdoor station FR005 (weekend correspond to 05/10 and 6/10).**

An important aspect to consider, in the analysis of the BS traffic, is the behavior of voice and data traffic over 24 hours: the former is characterized by very low traffic at nighttime, with an increase in the early hours of the morning, and two peaks corresponding to late morning and evening; the latter shows a similar trend, but with a phase shift of a few hours for the position of the traffic peaks.

In terms of percentage, moving from a subdivision of 60% voice and 40% data in the morning, 40% voice and 60% data in the evening, up to 10% voice and 90% data at night: the difference is certainly due to a different social behavior of end-users of the terminals (e.g. mobile phones, smartphones, tablets, etc.) during the day.

Data analysis shows that the difference in energy consumption between the periods of minimum and maximum traffic is about 400 Wh (in Fig. 2, from 2.4 kWh to 2.8 kWh). This behavior is also present in other types of BSs (shelter and room), though less prominent due to the energy contribution related to climate control. Moreover, it is quite clear that the energy consumption fluctuates daily along the traffic load trends. However, it should be emphasized that in the days when the station receives less telephone traffic (weekend), energy consumption does not change: indicating that there is no active energy management system, based on the traffic load.

A key concept to reduce the energy consumption in transmission is to turn off or on, depending on the needs of the traffic, the transceivers (TRXs) system of a BS. During periods of low traffic, TRXs, or also entire cells, can be placed in standby mode, thus avoiding the unnecessary waste of energy used to keep systems always on. This operation is usually called "power saving" [16][18][19][20][21].

In order to study the amount of energy that can be saved through a possible TRX shutdown, an algorithm based on the Monte Carlo method has been developed [22], which allows to simulate the behavior of a BS in "power saving" mode with different traffic loads (number and length of calls), characteristics and types (GSM, DCS); moreover, the algorithm also takes into account the Grade of Service of the BS, defined as the percentage of incoming calls to the BS that can be correctly handled, and simulates the TRX shutdown only if this does not affect the parameters of the station.

In the Monte Carlo simulations, input data coming from real traffic data is used. The number of new incoming calls has been supposed to have a Poisson distribution, while the length of the calls has been simulated by an Exponential distribution. The algorithm thus provides for every cycle a number of new calls and a call length for every call. The allocation of the calls in the available time slots is casual in the normal (No Power Saving) regime. Differently, in the Power Saving regime they are allocated in the free channel in the lower order TRX.

A TRX is switched OFF if it is left unused for a suitable number of cycles.

The procedure is iterated for the next cycle until the end of the simulation period, usually one day.

The power consumed by the BTS is calculated by summing the power consumed by the fixed apparatus (65 W), the power consumed by every channel in the control transmitter (12 W/channel), and the power consumed by every channel in the remaining active transmitters (9W/channel). These power values have been provided by the suppliers.

Fig. 3 shows the results of simulations performed on a station (FR001), from the real data traffic of a week (from 11 to 18 September 2013). This study analyzed the DCS technology with a TRX of 14 (446), maximum voice traffic of 80 erl on weekdays and 50/40 erlang at the weekend.

The abscissa in Fig. 3 indicates the day, the ordinate on the left represents the traffic load (erl) and the ordinate on the right is the percentage of energy saved. The blue line represents the weekly traffic variation, while the green line is the energy saved, obtained from the simulation with the Monte Carlo algorithm.

The energy saving is related to the amount of traffic (with less traffic, it is possible to turn off more TRX) and technical characteristic of BS. The station FR001 reports saving about 15% on weekdays, 20% on Saturday (September 14) and 24 % on Sunday (September 15).

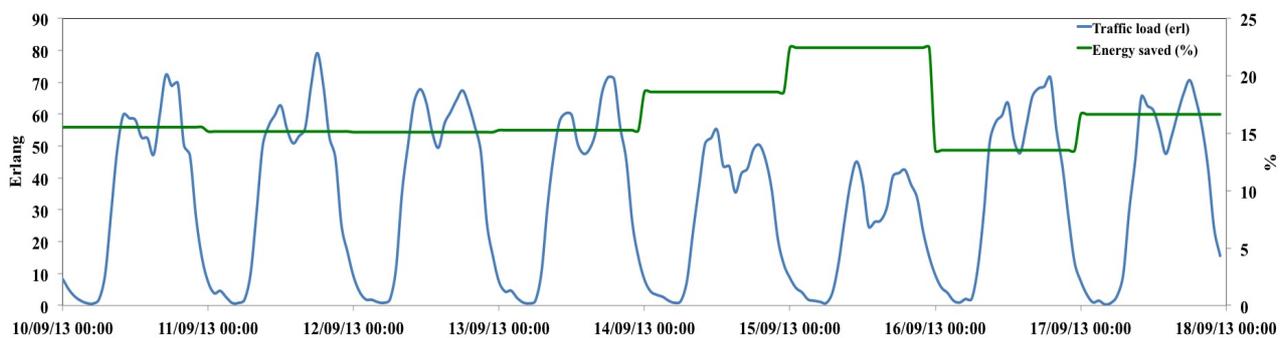

**Fig. 3. Simulated energy saving related to traffic load.**

## 4.3 Energy consumption versus air conditioning functions

The following stage of the study was the energy consumption trend, in relation to the need for air conditioning of the devices containing the equipment for the operation of the station. As mentioned in the previous paragraph, more than 40% of the energy needs of the shelters and the rooms are related to air conditioning.

For this purpose, the researcher measured the energy consumption and some environmental parameters and operation of the BS type "Shelter" and "Room", with the methods described above.

The "shelter" stations studied are the FR001 and FR009 (see Table 1) with the same number of TRXs and the same type of air-conditioning equipment and transmission. The two stations denoted a similar behavior. Below is the report and discussion of the results for the single station FR001.

In Fig. 4a-4d are the daily trends (for the day September 27, 2013) for energy consumption, global radiation, internal and external temperature and noise. It is possible to note for energy consumption (Fig. 4a), internal temperature (Fig. 4c) and noise (Fig. 4d), three distinct behaviors for three different periods of the day: 1) from midnight to about 9 AM, 2) from 9 AM to 6 PM, 3) from 6 PM to midnight.

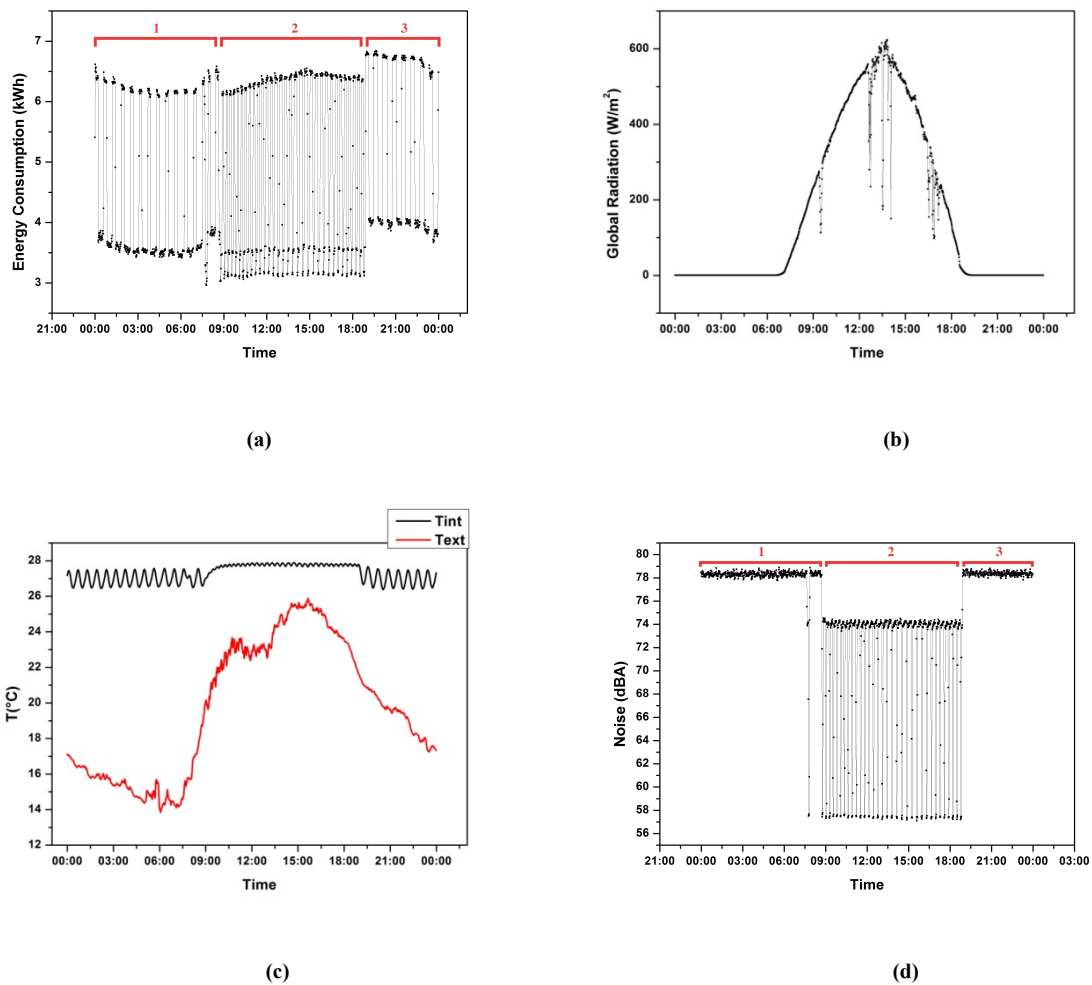

**Fig. 4. Daily trend for (a) energy consumption; (b) global radiation; (c) internal and external shelter temperature; (d) noise of the station FR001.**

Keep in mind that the day shown in the graphs is one of the hottest days of the sampling period; in fact, the external temperature (Fig. 4c, red line) reached a maximum value during the day of 26 °C.

The three time lapses will be analyzed as three distinct trends.

In the first (night) and the third period (evening), characterized by the lowest temperature of the day (Fig. 4c), with a minimum of 14 °C at night, energy consumption grew from about 3,5 kWh, from the free cooling systems (always ON) and the contribution of telephone traffic, to about 6.5 kWh due to the activation of the air conditioner 1.

The constant trend of the noise in the graph (Fig. 4d) is caused by the shelter extractor, which starts running at the activation of the free cooling, to prevent over-pressurization of the room. The internal temperature fluctuates around 27 degrees (temperature threshold of CDZ1).

Something different needs to be done for the period corresponding to the hottest hours of the day. In this case, the free cooling are off, since it is useless to extract hot air from the outside to cool the inside; hence, only the air conditioner is used. The energy consumption ranges from about 3 kWh (due to power systems of the entire BS added to the contribution of telephone traffic) to about 6 kWh when the air conditioning is turned on. Concerning the measurement of noise, the absence of an active free cooling causes the trend to oscillate and follow the activation-deactivation of the air conditioner.

The oscillation frequencies of energy consumption and internal temperature are related to the effects of hysteresis of the air conditioning systems (FC and CDZ) and/or to their operation cycles.

The overall trend (curvilinear) of energy consumption, with the highest values in the evening and the lowest at night and day, is also connected to the incidence of the traffic load on the latter.

Ultimately, it is possible to schematize the contributions of traffic and air-conditioning energy consumption as in Fig. 5, where POWER is the energy base demand of the BS; the terms En_FC1 and En_FC2 are the energy contribution coming from free cooling 1 and 2 respectively; En_CDZ1 and En_CDZ2 are the energy contribution coming from air conditioner 1 and 2 respectively; TRAFFIC is the energy contribution coming from the load traffic (during night period this contribution is almost irrelevant), and MIN and MAX correspond to the highest and the lowest value of the energy consumption in Fig. 4a.

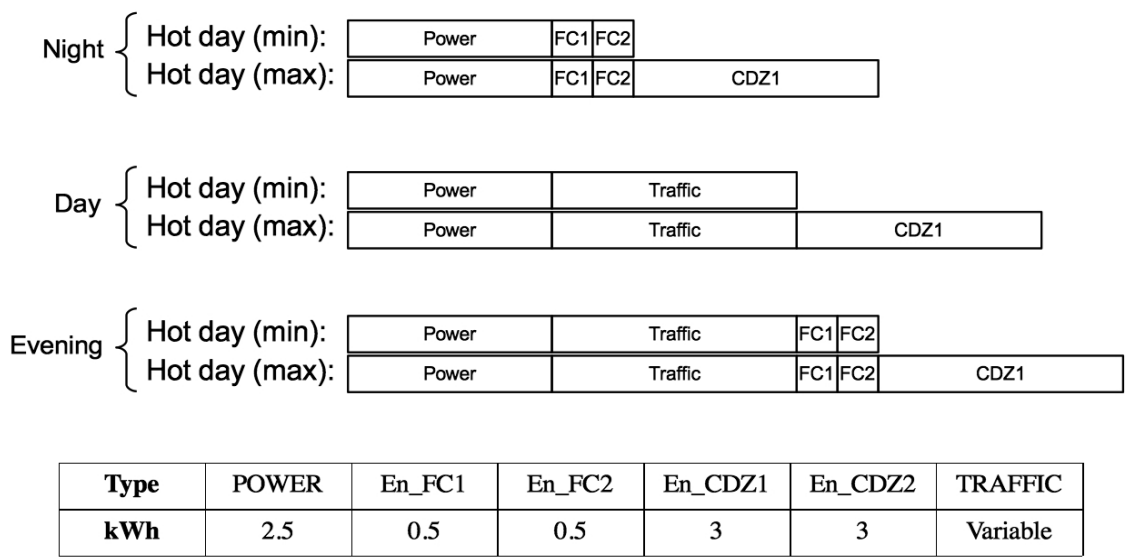

| Type | POWER | En_FC1 | En_FC2 | En_CDZ1 | En_CDZ2 | TRAFFIC |
|------|-------|--------|--------|---------|---------|---------|
| kWh  | 2.5   | 0.5    | 0.5    | 3       | 3       | Variable |

**Fig. 5. Contributions of traffic load and air conditioning system energy consumption related to period (night, day, evening).**

The situation is different in the case of cold days (e.g. December 7, 2013 with the maximum temperature equal to 13 °C). During the night and in the morning, power is consumed only by the free cooling, since the cold air from the outside is sufficient to lower the temperature inside the shelter. When the internal temperature begins to increase, due to the outdoor temperature, one of the air conditioners starts.

In this case, there is a clearly different energy consumption behavior. In fact, it is lower than during a hot day, because even during the day, the two conditioners are never turned on because the external temperature is lower and the free cooling can cool the internal part.

The two BSs type "Room" have the same number of TRXs and the same type of conditioning and transmission devices. The behavior is the same as the "shelter" stations, where the contribution of the air-conditioning equipment is clear. In fact, the temperature rises until it reaches the threshold for activation of the air conditioner, and then moves to lower values. This allows turning off the air conditioners and leaving the two free cooling ON.

A factor to keep under control in "Outdoor" BS is the high internal temperatures reached by the device. Fig. 6 shows the values of the temperature outside and inside the box that contains the transmission equipment for the BS FR005. It is clear that the internal temperature depends on the external temperature; moreover, even for moderately high external temperatures, the cabinet will reach a high internal temperature; for example, in the specific case, it comes to over 45 °C, although the outside temperature is not higher than 20 °C.

Please note that the dependence between outside and inside temperature is lost in the BSs type shelter and room, when the free-cooling and air-conditioning function are activated based on on-off logic.

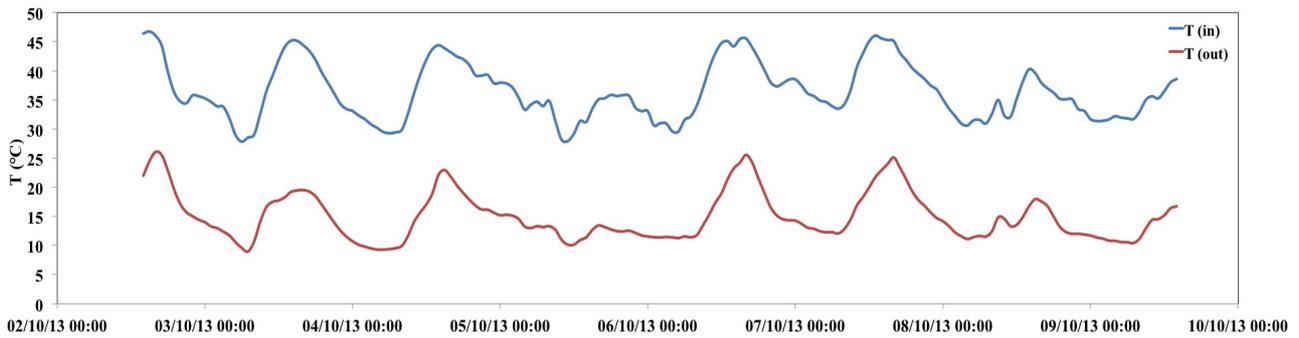

**Fig. 6. Internal and external temperature trends for the station FR005.**

**5. Metrics and standardization**

For the evaluation of energy performance of a BS, it is useful to identify the metrics that allow to compare energy consumption regardless of their operating characteristics (type, technology, etc.) and their location (urban, rural, etc.).
This standardization work is being carried out by all the suppliers and international service providers through their associations, such as the ITU and ETSI, which have set up special committees to promote and produce their own standards to be used globally in the field of telecommunications. In particular, they are developing harmonized methods to assess the energy efficiency of BS through universal indicators [23].
Here is an example of two possible indicators:
**Indicator I1: Traffic/ Energy Consumption (erl/kWh – Bit/kWh)**
**Indicator I2: Coverage Area/ Energy Consumption ($m^2$/ kWh)**
There is an ongoing discussion as to whether these indicators can actually allow to compare energy performance of BSs that differ in type and technology.
Essential for the calculation of these metrics are the following parameters:
- **Timing advance (TA):** time required for a signal to reach the radio base station from a mobile phone. Each unit of TA corresponds to 550m.
- **Measurement report (MR):** data samples sent by the mobile station every 480ms, regarding the change of the network and radio parameters. The MR can be used as an estimate of the users (mobile stations) that are within range of the station.

Fig. 7 outlines the performance of the measurement report as a function of the timing advance, for station FR001. It is easy to notice the decrease of the measurement reports, the larger the distance from the station.
As can be seen, 90 % of MR is located in the range of 10 TA, and this will be the limit value of the calculation, entailing a coverage of the BS at a distance of 5500 meters.

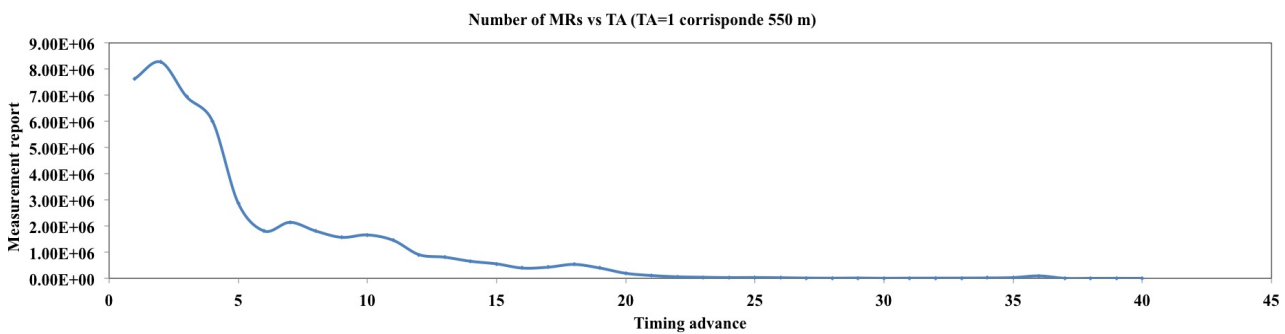

**Fig. 7. Measurement report vs timing advance.**

As reported in the previous paragraphs, the average energy consumption of FR001 is 4.86 kWh, with a total annual consumption of 42,705 kWh, while the FR005 average consumption is 2.7 kWh, with a total annual consumption of 23,725 kWh. These values allow to calculate the energy consumption per telephone traffic load (Indicator I1) and per surface (Indicator I2), according to the ETSI metrics for the BS monitored.
In the following has been calculated the indicators I1 and I2 for the stations FR001 and FR005.

**FR001 -** Indicator I1 (per day): 104.18/116.55 erl/kWh = 0.89 erl/kWh
**FR005 -** Indicator I1 (per day): 60.26/64.75 erl/kWh = 0.93 erl/kWh
**FR001 -** Indicator I2 (TA=10, annual consumption):  2,231 $m^2$/ kWh
**FR005 -** Indicator I2 (TA=7, annual consumption):  2,000 $m^2$/ kWh

**6. Conclusion**

Energy monitoring was carried out on 6 BSs located in center Italy and belonging to 3 different types (shelter, room, outdoor). According to the type of BS, the power consumption is greater in shelter and room than outdoors; in this particular case ranging from an average of 39,420 kWh/year for the shelter, to 27,923 kWh/year for room and 19,345 kWh/year for the outdoors. This difference in consumption is mainly due to the need of shelter and room to use air conditioning systems inside the building.
Energy consumption follow the traffic load, even if they does not change when the station has less telephone traffic (days of weekend), i.e. the energy management of the radio base station is independent of the telephone traffic load. Moreover, it has been emphasized that the voice traffic and data traffic patterns are slightly different over 24h reflecting the different "social behavior" of the end-users.
A specific "TRX intelligent shutdown" simulation algorithm has been used to show that an energy saving of about 20% can be obtained.
Measurements show that air conditioning systems cover the 40 – 45 % of the total energy consumption. The principal behavior of the "Shelter air conditioning systems" has been described as the functions of the single apparatus. For typology "BS outdoor" the temperatures of the apparatuses has been monitored, founding very high temperatures even with external mild temperatures.
In order to make a comparison between stations with different characteristics and localizations, a standardization process is needed. An indicator, proposed by ETSI and ITU, has been used for the BSs monitored.

**Acknowledgements.**

We kindly acknowledge Italian mobile telecommunications providers Wind for the use of experimental sites. This study was partially funded by Regione Campania through European funds (POR Campania FSE 2007-2013) for the "Dottorato in azienda" project.